\documentclass[%
 reprint,
superscriptaddress,
preprint,
onecolumn,
 amsmath,amssymb,
 aps,
prb,
]{revtex4-1}

\usepackage{graphicx}
\usepackage{dcolumn}
\usepackage{bm}

\begin{document}


\title{Influence of Helical Spin Structure on the Magnetoresistance of an Ideal Topological Insulator}

\author{T. Ozturk}
\email{teozturk@gmail.com}
\affiliation{Department of Physics, University of Michigan, Ann Arbor, MI 48109, USA.}
\affiliation{Department of Physics, Selcuk University, Konya, 42075, Turkey.}
\author{R. L. Field III}
\affiliation{Department of Physics, University of Michigan, Ann Arbor, MI 48109, USA.}
\author{Y. S. Eo}
\affiliation{Department of Physics, University of Michigan, Ann Arbor, MI 48109, USA.}
\author{S. Wolgast}
\affiliation{Department of Physics, University of Michigan, Ann Arbor, MI 48109, USA.}
\author{K. Sun}
\affiliation{Department of Physics, University of Michigan, Ann Arbor, MI 48109, USA.}
\author{C. Kurdak}
\affiliation{Department of Physics, University of Michigan, Ann Arbor, MI 48109, USA.}


\date{\today}

\begin{abstract}
In an ideal topological insulator, the helical spin structure of surface electrons suppresses backscattering and thus can enhance surface conductivity. We investigate the effect of perpendicular magnetic field on the spin structure of electrons at the Fermi energy and calculate a magnetic-field dependent topological enhancement factor for different disorder potentials, ranging from short-range disorder to screened Coulomb potential. Within the Boltzmann approximation, the topological enhancement factor reaches its maximum value of 4 for a short-range disorder at zero magnetic field and approaches a value of 1 at high magnetic fields independent of the nature of the disorder potential.  
\end{abstract}

\maketitle


\section{Introduction}

A topological insulator (TI) is a new class of material with a bandgap in the bulk and gapless surface states arising from the topology of the bulk quantum wavefunctions \cite{RevModPhys.82.3045,Moore2010,RevModPhys.83.1057}. The surface states are protected by time-reversal symmetry and are expected to dominate conduction at low temperatures. TIs were first predicted in two-dimensional semiconductors \cite{PhysRevLett.95.226801,PhysRevLett.95.146802,Bernevig15122006} and later generalized to three-dimensional systems \cite{PhysRevLett.98.106803,PhysRevB.75.121306}, such as Bi$_{1-x}$Sb$_{x}$, Bi$_{2}$Se$_{3}$, Bi$_{2}$Te$_{3}$ and Sb$_{2}$Te$_{3}$ \cite{PhysRevB.76.045302,Murakami2007,Hsieh2008,Zhang2009,Xia2009,Chen10072009}. The presence of topologically protected surface states was first tested using angle-resolved photoemission spectroscopy (ARPES) measurements, which revealed a linear dispersion for these states \cite{Hsieh2008,Xia2009,Chen10072009}. In addition to the linear dispersion, the TIs must exhibit a helical spin structure, which is also confirmed by spin-resolved ARPES measurements \cite{Roushan2009,Hsieh2009}. However, most experimentally studied TIs such as Bi$_{1-x}$Sb$_{x}$ \cite{PhysRevB.80.085303} and Bi$_{2}$Te$_{2}$Se \cite{PhysRevB.82.241306,Xiong2012} exhibit surface conduction which is polluted by defect-assisted bulk transport, thereby presenting challenges to separating the bulk from surface conduction. In contrast, for SmB$_6$, recent experiments reveal that the surface dominant regime has been achieved \cite{PhysRevB.88.180405,Kim2013}.

The helical spin structure of TIs induces unique transport properties. Most significantly, the locking of spin to the momentum direction inhibits backscattering in the absence of the external magnetic field \cite{Roushan2009,PhysRevLett.103.266803} and this leads to a quantum correction to conductivity, known as weak anti-localization \cite{PhysRevLett.106.166805,PhysRevLett.108.036805}. Also, compared to the classical framework of conductivity for normal metals, the locking of the spins for TIs results in an enhancement of conductivity. Although this enhancement is well-known to the community, the unique response to external magnetic fields is less understood. The role of spin-locking must be theoretically resolved to aid recent experimental magnetoconductivity reports on TIs. Here, we calculate the magnetic field dependence of an enhancement factor to conductivity for different types of disorder potentials. We consider the effects of perpendicular magnetic field on the spin structure of surface states for an ideal TI. We find that the helical spin structure leads to an additional positive magnetoresistance at large magnetic fields, which is perpendicular to the surface-conducting layer. Furthermore, the size of the magnetoresistance is found to be greatest for short-range disorder.      

\section{HELICAL SPIN STRUCTURE}

\begin{figure}[ht]
\includegraphics[width=8.6cm]{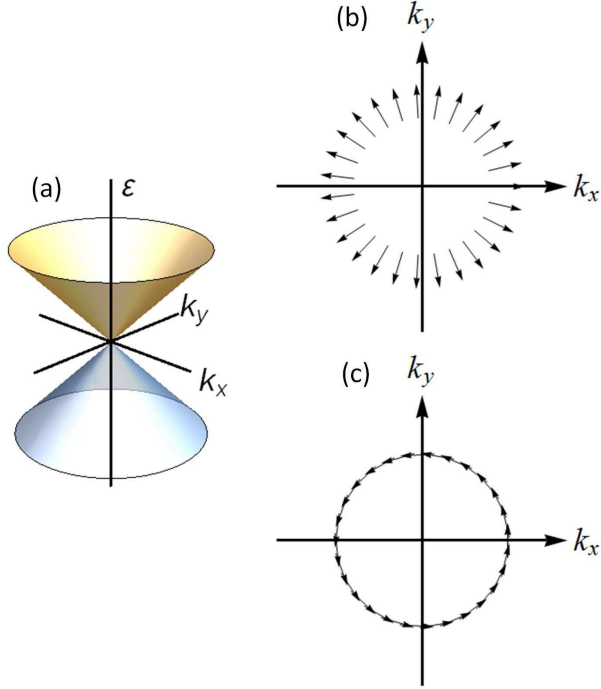}
\caption{\label{fig_1} (a) Energy versus wave number dispersion of an ideal TI. (b)-(c) Schematic illustrations of two possible helical spin structures where the spin is parallel or perpendicular to the wave number, respectively.}
\end{figure}

The Hamiltonian of the surface states of an ideal TI is well known in two forms. Considering the surface perpendicular to the $\hat{\bm{z}}$-direction, the Hamiltonians describing the low-energy physics of the surface states can be expressed as \cite{Zhang2009,PhysRevB.85.235413}:
\begin{equation}
\label{eq:one}
H_{1}={\hbar}v_{F}\bm{\sigma}\cdot\bm{k}
\end{equation}
\begin{equation}
\label{eq:two}
H_{2}={\hbar}v_{F}\left(\bm{\sigma}\times\hat{\bm{z}}\right)\cdot\bm{k}
\end{equation}     
where $v_{F}$ is the Fermi velocity, $\bm{\sigma}$ denotes the Pauli spin matrices and $\bm{k}=\left(k_{x},k_{y}\right)$ is the wave vector. The eigenvalues of these surface Hamiltonians are given by
\begin{equation}
\label{eq:three}
\varepsilon=\pm{\hbar}v_{F}\sqrt{\left(k_x^2+k_y^2\right)}=\pm{\hbar}v_{F}k
\end{equation}
leading to surface states with a Dirac cone type dispersion, as shown schematically in Fig.~\ref{fig_1}(a). Here we note that for real 3D TIs, this linear dispersion holds only near the Dirac point. The linear dispersion bends by the intrinsic nature of the bulk. Furthermore, the eigenvectors of these two Hamiltonians are given by 
\begin{equation}
\label{eq:four}
\Psi_{1\pm}=\frac{1}{\sqrt{2}}\begin{pmatrix} {\pm}e^{-i\phi}\\1 \end{pmatrix}
\end{equation}
\begin{equation}
\label{eq:five}
\Psi_{2\pm}=\frac{1}{\sqrt{2}}\begin{pmatrix} {\mp}ie^{-i\phi}\\1 \end{pmatrix}
\end{equation}
where $\phi$ is the angle of $\bm{k}$ with respect to $k_{x}$. In both cases the eigenvectors indicate a helical spin structure where the spin direction is tied to the direction of the wave vectors. For the first (second) case the spin direction is parallel (perpendicular) to the wave vectors, as illustrated in Fig.~\ref{fig_1}(b) and (c).

Our goal is to include the helical spin structure in the calculation of scattering rates. Typically the spin structure leads to a suppression of scattering rates, which is dependent on the angle of scattering. At zero magnetic field, the total backscattering of electrons is not allowed (from the momentum $\bm{k}$ to $\bm{-k}$). We should note that since the overall spin angles between the two Hamiltonians at the same momentum consistently differ by 90 degrees, the suppression factors are identical for both Hamiltonians; here, we will only focus on the first Hamiltonian for the rest of this paper.

In the presence of an external magnetic field, the surface Hamiltonian of a TI will have an additional Zeeman term:
\begin{equation}
\label{eq:six}
H_{total}={\hbar}v_{F}\bm{\sigma}\cdot\bm{k}+g{\mu_{B}}\bm{\sigma}\cdot\bm{B}
\end{equation}
where $\bm{B}$ is an external magnetic field, $g$ is the Lande-g factor and  $\mu_{B}$ is the Bohr magneton. In addition to the Zeeman splitting, the magnetic field also couples to the orbital motion (e.g. via minimal coupling), which plays a crucial role in the understanding of quantum phenomena like quantum interferences and the formation of Landau levels. Here, for simplicity, we will only focus on the Zeeman coupling while the minimal coupling is ignored. This approximation is justified for field parallel to the surface, where the minimal coupling drops out for a 2D system. However, for perpendicular magnetic fields, additional quantum corrections may arise, which are beyond the scope of this investigation. We will first consider the influence of a parallel magnetic field. The dispersion relation is shifted in the presence of a parallel magnetic field: 
\begin{equation}
\label{eq:seven}
\varepsilon_{\parallel}=\pm\sqrt{\left({\hbar}v_{F}k_{x}+g{\mu_{B}}B_{x}\right)^2+\left({\hbar}v_{F}k_{y}+g{\mu_{B}}B_{y}\right)^2}
\end{equation}
Thus, for a given $\bm{k}$, the presence of an external magnetic field alters the direction of the spin. However, if we redefine a new wave number with respect to the position of the Dirac cone, the spin direction of the electron for the shifted wave number would be the same as that of the case of zero magnetic field. For example, the spin structure at the Fermi energy will be identical to the helical spin structure illustrated in Fig.~\ref{fig_1}(b). Since, the magnetic field only shifts the position of the Dirac cone without changing the spin structure and the scattering rates would not dependent on the magnitude of parallel magnetic field. And therefore, the helical spin structure does not play a role in magnetoresistance in a parallel magnetic field.

\begin{figure}[ht]
\includegraphics[width=8.6cm]{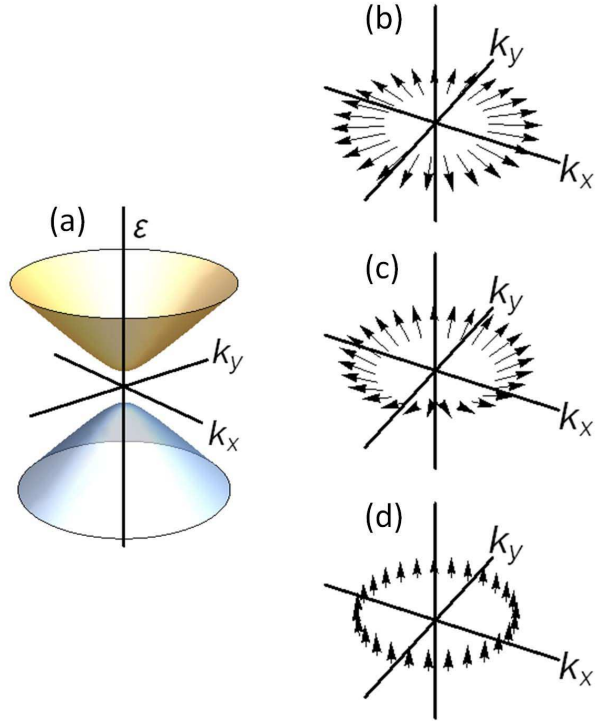}
\caption{\label{fig_2} (a) Energy versus wave number dispersion of an ideal TI in perpendicular magnetic field. (b), (c), and (d) The helical spin structures of electrons at a constant energy circle are illustrated schematically for different values of magnetic field: $B=0$, $B=B_{TI}$, and $B=\infty$ , respectively.}
\end{figure}

Now let us consider the influence of a perpendicular magnetic field. Unlike the case of parallel magnetic field, the Dirac cone does not survive, as is evident in the modified dispersion relation
\begin{equation}
\label{eq:eight}
\varepsilon_{\perp}=\pm\sqrt{\left(g{\mu_{B}}B_{z}\right)^2+\left({\hbar}v_{F}k_{x}\right)^2+\left({\hbar}v_{F}k_{y}\right)^2}
\end{equation}
Most significantly, there is an opening of a gap with an energy of $2g{\mu_{B}}B_{z}$ at $k=0$. The energy dispersion in Eq.~(\ref{eq:eight}) is shown in Fig.~\ref{fig_2}(a). We also calculate the eigenvectors which are given as
\begin{equation}
\label{eq:nine}
\Psi_{\perp+}=\begin{pmatrix} -\cos\frac{\eta}{2}e^{-i\phi}\\-\sin\frac{\eta}{2} \end{pmatrix},\Psi_{\perp-}=\begin{pmatrix} \sin\frac{\eta}{2}e^{-i\phi}\\-\cos\frac{\eta}{2} \end{pmatrix}
\end{equation}
where $\eta$ is 
\begin{equation}
\label{eq:ten}
\eta=\tan^{-1}\left(\frac{{\hbar}v_{F}k}{g{\mu_{B}}B_{z}}\right).
\end{equation}
We should note that $\eta$ has a geometric interpretation. It is instructive to rewrite the Hamiltonian in Eq.~(\ref{eq:six}) as
\begin{equation}
\label{eq:eleven}
H_{total}=g{\mu_{B}}\left[\bm{\sigma}\cdot\left(\frac{{\hbar}v_{F}\bm{k}}{g{\mu_{B}}}+\bm{B}\right)\right]
\end{equation}
where the quantity $\frac{{\hbar}v_{F}\bm{k}}{g{\mu_{B}}}$ acts as an intrinsic magnetic field ($\bm{B}_{TI}$) of TI. Then the effective magnetic field becomes the sum of these two fields:
\begin{equation}
\label{eq:twelve}
\bm{B}_{eff}=\frac{{\hbar}v_{F}\bm{k}}{g{\mu_{B}}}+\bm{B}
\end{equation}
When written this way it is clear that spin direction is determined by $\bm{B}_{eff}$. The $\eta$, which is defined in Eq.~(\ref{eq:ten}), is the angle of $\bm{B}_{eff}$ with respect to the $z$-direction. The first term of the above equation acts as an intrinsic magnetic field. At the Fermi energy, the magnitude of this intrinsic magnetic field is written as
\begin{equation}
\label{eq:thirteen}
B_{TI}=\frac{{\hbar}v_{F}k_F}{g{\mu_{B}}}
\end{equation}
The spin structure at the Fermi energy depends on the ratio of perpendicular magnetic field to the magnitude of intrinsic magnetic field of the TI. The spin structures for different values of $B/B_{TI}$ are schematically shown in Fig.~\ref{fig_2}(b)-2(d): With the increasing magnitude of the perpendicular magnetic field the spins tilt towards the $z$-direction and at $B=\infty$ (shown in Fig.~\ref{fig_2}(d)) the spins would be fully aligned. From the perspective of transport we would expect a positive magnetoresistance and for the backscattering to be enhanced as the magnitude of perpendicular magnetic field increases.

\section{TOPOLOGICAL ENHANCEMENT FACTOR}

In order to understand the influence of the helical spin structure on transport properties we need to focus on how its spin structure changes the momentum relaxation time. For an ordinary two-dimensional conductor the momentum relaxation time can be written using the Boltzmann transport equation as
\begin{equation}
\label{eq:fourteen}
\frac{1}{\tau}=\int W\left(k,k^{\prime}\right)\left(1-\cos{\theta}\right)d^{2}k
\end{equation}
where $\theta$ is the angle between $\bm{k}$ and $\bm{k}^{\prime}$ and $W\left(k,k^{\prime}\right)$ is the scattering rate of these two wave vectors. For a given disorder potential, the scattering rate can be calculated using Fermi's Golden Rule. In the case of a TI there would be an addition suppression factor in the transport equation arising from the helical spin structure:
\begin{equation}
\label{eq:fifteen}
\frac{1}{\tau_{TI}}=\int W\left(k,k^{\prime}\right)S\left(s,s^{\prime}\right)\left(1-\cos{\theta}\right)d^{2}k,
\end{equation}
where $s$ ($s^{\prime}$) is the spin of an electron with a wave number $k$ ($k^{\prime}$) and  $S\left(s,s^{\prime}\right)$ is the contribution to the scattering rate due to the spin structure of TI. We note that spin factor $S\left(s,s^{\prime}\right)$ is 1 when $s$ and $s^{\prime}$ are parallel to each other and 0 when $s$ and $s^{\prime}$ are antiparallel.
 
For an isotropic conductor the scattering rate and the spin factor would only depend on the angle between $\bm{k}$ and $\bm{k}^{\prime}$, also known as the scattering angle $\theta$: $W\left(k,k^{\prime}\right)=W\left(\theta\right)$ and  $S\left(s,s^{\prime}\right)=S(\theta)$. Thus, $\theta$ becomes the integration variable for the above integrals. Since the spin factor $S\left(s,s^{\prime}\right)$ is bounded by 0 and 1, the momentum relaxation time of a TI, $\tau_{TI}$, is always greater than that of an identical electron system without a spin structure, $\tau$.  To quantify the contribution of the helical spin structure to transport we introduce a topological enhancement factor, $\Xi=\frac{\tau_{TI}}{\tau}$, which is defined as the ratio of these two momentum relaxation times. We expect $\Xi$ would be dependent on the nature of the disorder potential as well as the applied magnetic field. 

In the absence of external magnetic field, the spin contribution to the scattering rate, $S\left(s,s^{\prime}\right)$, can be written as
\begin{equation}
\label{eq:sixteen}
S\left(s,s^{\prime}\right)=\cos^2\left(\frac{\theta}{2}\right)
\end{equation}
Thus, as expected at $\theta=\pi$, the spin dependent scattering rate will be zero and subsequently there is no backscattering. In the presence of an external magnetic field in the $z$-direction, spins are expected to be parallel to the effective magnetic field and can be written as:
\begin{equation}
\label{eq:seventeen}
\bm{s}=\frac{1}{\sqrt{B_{TI}^2+B_{z}^2}}\left(B_{TI}\cos{\phi},B_{TI}\sin{\phi},B_{z}\right)
\end{equation}
\begin{equation}
\label{eq:eighteen}
\bm{s}^\prime=\frac{1}{\sqrt{B_{TI}^2+B_{z}^2}}\left(B_{TI}\cos\left(\theta+\phi\right),B_{TI}\sin\left(\theta+\phi\right),B_{z}\right)
\end{equation}
Defining the angle between these two spins as $\theta_{ss^\prime}=\cos^{-1}\left(\frac{\bm{s}\bm{s}^\prime}{\left|\bm{s}\right|\left|\bm{s}^\prime\right|}\right)$, the spin contribution to the scattering rate $S\left(s,s^{\prime}\right)$ becomes
\begin{equation}
\label{eq:nineteen}
S\left(s,s^{\prime}\right)=\cos^2\left(\frac{\theta_{ss^\prime}}{2}\right)=\cos^2\left[\frac{1}{2}\cos^{-1}\left(\frac{B_{TI}^2\cos{\theta}+B_{z}^2}{B_{TI}^2+B_{z}^2}\right)\right]
\end{equation}
Knowing $S\left(s,s^{\prime}\right)$ we can now calculate $\Xi$ for different disorder potentials. 

In this paper, we only discuss the magnetic response to $\Xi$ for short-range and long-range disorder potentials. We note that another scattering mechanism of great interest \cite{Cha2010} is magnetic impurities which are not included in this work. Even for the simplified mean-field approach, a single magnetic impurity additionally deforms the helical spin angles, which have only been calculated numerically \cite{PhysRevLett.102.156603}. Spin-flip scattering by magnetic impurities, which gives rise to the Kondo effect, can be even more demanding theoretically \cite{Xin2013}. 

\subsection{Short-Range Disorder Potential}

Let us start with the simplest case where the disorder potential is short-range. Examples of short-range disorder include alloy disorder and surface or interface roughness. In all these cases the scattering rate is independent of the scattering angle. Thus, the disorder potential is characterized by a single parameter, $W\left(k,k^{\prime}\right)=W_{0}$ and the calculation of $\Xi\left(B\right)$ can be performed analytically. We then obtain an analytical solution for $\Xi\left(B\right)$, considering the magnetic-field-dependent spin contribution for a constant scattering rate:
\begin{equation}
\label{eq:twenty}
\Xi\left(B\right)=\frac{\displaystyle \int W_{0}\left(1-\cos{\theta}\right)d^{2}k}{\displaystyle \int W_{0}\cos^2\left[\frac{1}{2}\cos^{-1}\left(\frac{B_{TI}^2\cos{\theta}+B_{z}^2}{B_{TI}^2+B_{z}^2}\right)\right]\left(1-\cos{\theta}\right)d^{2}k}=\frac{4\left(B_{TI}^2+B_{z}^2\right)}{B_{TI}^2+4B_{z}^2}
\end{equation}
As expected the topological enhancement factor decreases from 4 to 1 smoothly with increasing magnetic field.  In the low magnetic field limit, there is a quadratic correction to magnetoconductivity due to $\Xi\left(B\right)$. 

\subsection{Long-range disorder potential}

\begin{figure}[ht]
\includegraphics[width=8.6cm]{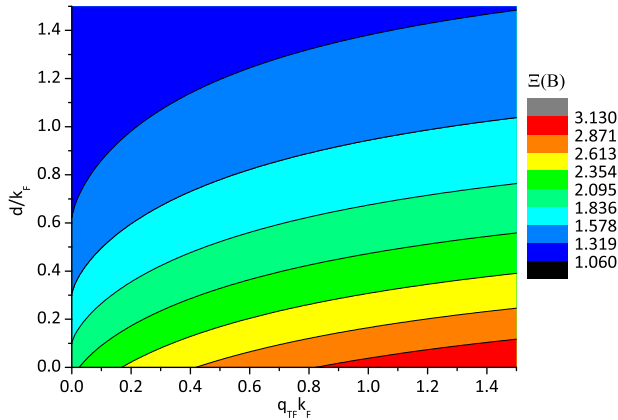}
\caption{\label{fig_3} Contour graph of the topological enhancement factors for different $d/k_F$ distances and $q_{TF}k_F$ values.}
\end{figure}

We now consider the case of the long-range Coulomb disorder caused by charged impurities. Unless the interfaces are protected by the growth of an epitaxial heterostructure, one would expect a large concentration of charged impurities in close vicinity of the two-dimensional topological surface states. In order to write down the disorder potential we need to know the dielectric constant of the material as well as the location of charged impurities. In the calculation below, we will focus on a single impurity located a distance $d$ from the two-dimensional electron system. In this case, the scattering rate can be written as \cite{PhysRevB.82.155457} 
\begin{equation}
\label{eq:twentyone}
W\left(k,k^{\prime}\right)\propto \left|\frac{U\left(q\right)}{\epsilon\left(q\right)}\right|^2
\end{equation}
where, $q=\left|\bm{k}-\bm{k}^{\prime}\right|=2{k_F}\sin\frac{\theta}{2}$ and $U\left(q\right)\propto e^{-qd}/\left(q\right)$ is a screened Coulomb potential in Fourier domain \cite{PhysRevLett.98.186806}. The screening of the Coulomb potential by the two-dimensional electron system can be included in the calculation by the use of the following dielectric function
\begin{equation}
\label{eq:twentytwo}
\epsilon\left(q\right)=\epsilon_{0}\left(1+\frac{q_{TF}}{q}\right)
\end{equation}
where $q_{TF}$ is given by the inverse of the Thomas-Fermi (TF) screening length. Then, the scattering rate can be written as
\begin{equation}
\label{eq:twentythree}
W\left(\theta\right)\propto\frac{e^{-2d\left|q\right|}}{\left(q+q_{TF}\right)^2}
\end{equation}
Thus, for Coulomb scattering the topological enhancement factor is expected to depend on both $q_{TF}$ and $d$, which can be calculated numerically. A contour plot of $\Xi$ calculated  at zero magnetic field is shown in Fig.~\ref{fig_3}. The topological enhancement factor for Coulomb scattering is smaller than that for short range disorder.

\begin{figure}[ht]
\includegraphics[width=8.6cm]{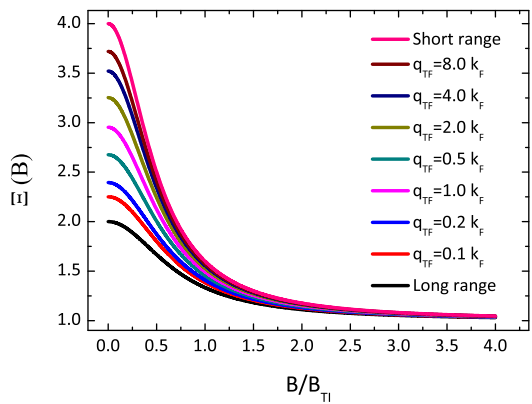}
\caption{\label{fig_4} The topological enhancement factors according to $B/B_{TI}$ for short-range potential, long-range potential and for different $q_{TF}$ values.}
\end{figure}

We should note that there is a special case where $d=0$ and $q_{TF}=0$ for which the calculation can be performed analytically. In this special case, the charged impurities would be located at the 2DEG and there would be no screening and thus we can express the potential of a charged impurity as a simplified Coulomb potential \cite{PhysRevLett.98.186806}. Then, we can write the scattering rate for the impurities in the absence of magnetic field as
\begin{equation}
\label{eq:twentyfour}
W\left(k,k^{\prime}\right)=\frac{1}{q^2}
\end{equation}
Using this scattering rate along with the spin contribution in Eq.~(\ref{eq:nineteen}), for the contributions of charged impurities under the influence of an external magnetic field, we obtain an analytical solution for $\Xi\left(B\right)$:
\begin{equation}
\label{eq:twentyfive}
\Xi\left(B\right)=\frac{\displaystyle \int \frac{1}{q^2}\left(1-\cos{\theta}\right)d^{2}k}{\displaystyle \int \frac{1}{q^2}\cos^2\left[\frac{1}{2}\cos^{-1}\left(\frac{B_{TI}^2\cos{\theta}+B_{z}^2}{B_{TI}^2+B_{z}^2}\right)\right]\left(1-\cos{\theta}\right)d^{2}k}=\frac{2\left(B_{TI}^2+B_{z}^2\right)}{B_{TI}^2+2B_{z}^2}
\end{equation}
The results of the topological enhancement factor under magnetic field for both short-range and long-range disorder are illustrated in Fig.~\ref{fig_4}.  In this figure, we assume that $q_{TF}$ and $d$ are proportional to $k_F$, ranging from short-range constant scattering to long-range Coulomb potential. For intermediate cases between short-range and long-range potential, we assume TF screening theory (Eq.~(\ref{eq:twentythree})) assuming that $d=0$, which signifies that the impurity layer is very close to the surface. In the absence of magnetic field, $\Xi\left(B\right)$ increases from 2 to 4 as $q_{TF}$ increases from long-range to short-range potential. In all cases, $\Xi\left(B\right)$ approaches 1 as $B/B_{TI}$ increases. Thus, TF screening theory is consistent with the calculations from long-range and short-range potentials.

We should note that for most TIs, $B_{TI}$ is large, and thus, the corrections to magnetoconductivity are expected to be small. For example, Bi$_2$Se$_3$ \cite{Xia2009,PSSR:PSSR201206457} has a single Dirac cone with a Fermi velocity $v_F=5\times10^{5}$ ms$^{-1}$ and a Fermi wavevector $k_F\approx1$ nm$^{-1}$, leading to a $B_{TI}\approx2840$ T. Similarly, for the topological Kondo insulator SmB$_6$, Fermi velocity and Fermi wavevector have been measured for two different Fermi pockets \cite{Liarxiv}. The first Fermi pocket has a Fermi velocity of $\sim5\times{10}^{5}$  ms$^{-1}$ and Fermi wavevector of $k_F\approx0.383$ nm$^{-1}$, so $B_{TI}\approx1307$ T. For the second Fermi pocket, $v_F=10.9\times{10}^{5}$  ms$^{-1}$ and $k_F\approx0.955$ nm$^{-1}$, so $B_{TI}\approx5918$ T. Even in such high carrier density two-dimensional systems, the quadratic correction to magnetoconductivity due to $\Xi\left(B\right)$ can have a measurable effect and should be included in the analysis of high field data \cite{Wolgastarxiv}. On the other hand, if we can realize topological conducting states with a very low Fermi wavevector, we could reach a regime where $B/B_{TI}$ is large and there is a total suppression of the topological enhancement factor.  

\section{CONCLUSION}

In conclusion, we have studied theoretically the magnetoresistance of an ideal TI surface utilizing Boltzmann transport. We investigate conductivity enhancement in the presence of a perpendicular magnetic field. By finding the angle deflection of the helical spins by the Zeeman effect, the enhancement is calculated within the classical scattering theory. For short-range disorder in the absence of magnetic field, $\Xi\left(B\right)$ is equal to 4. With increasing magnetic field the topological enhancement factor decreases from 4 to 1 smoothly as given by the analytical function of Eq.~(\ref{eq:twenty}). For Coulomb disorder with no screening, the topological enhancement factor is equal to 2 in the absence of the magnetic field.  In the presence of magnetic field the topological enhancement factor is suppressed by the analytical function given in Eq.~(\ref{eq:twentythree}). We also investigated the effect of the screening using TF screening theory. The results are relevant for all high field magnetotransport measurements performed on TIs. The corrections to magnetoconductivity are small for high carrier density systems such as the conducting electrons on the surface of SmB$_{6}$, but are expected to be more significant for TIs with lower carrier density of electrons.

\begin{acknowledgments}
This work is supported by TUBITAK (The Scientific and Technological Research Council of Turkey) and by the National Science Foundation through DMR-1006500, DMR 1441965 and PHY 1402971. 
\end{acknowledgments}

\nocite{*}

\bibliography{references}

\end{document}